\documentclass[preprint]{aastex}

\newcommand{\tcltk}{{\it Tcl/tk}}
\newcommand{\C}{{\it C}}
\newcommand{\kms}{km~s$^{-1}$}
\newcommand{\um}{$\mu$m}
\newcommand{\apeq}{$\approx$}

\begin{document}

\title{TEXES: A Sensitive High-Resolution Grating Spectrograph for the
Mid-Infrared}
\author{J. H. Lacy, M. J. Richter, T. K. Greathouse, D. T. Jaffe, and Q. Zhu}
\affil{Astronomy Department, University of Texas, Austin, TX 78712}
\email{lacy@astro.as.utexas.edu, richter@astro.as.utexas.edu, 
tommyg@astro.as.utexas.edu, dtj@astro.as.utexas.edu, 
zhuqf@astro.as.utexas.edu}

\begin{abstract}
We discuss the design and performance of TEXES, the Texas Echelon Cross
Echelle Spectrograph.
TEXES is a mid-infrared (5-25~\um) spectrograph with several operating modes:
high-resolution, cross-dispersed with a resolving power of
R~=~$\lambda / \delta \lambda$~\apeq~100,000, 0.5\% spectral
coverage, and a $\sim 1.5 \times 8^{\prime\prime}$ slit;
medium-resolution long-slit, with R \apeq\ 15,000, 0.5\% coverage,
and a $\sim 1.5 \times 45$\arcsec\ slit;
low-resolution long-slit, with $\delta \lambda$~\apeq~.004~\um,
0.25~\um\ coverage, and a $\sim 1.5 \times 45$\arcsec\ slit;
and source acquisition imaging with 0.33\arcsec\ pixels and
a $25 \times 25$\arcsec\ field of view on a 3~m telescope.
TEXES has been used at the McDonald Observatory 2.7m and the IRTF 3m
telescopes, and has proven to be both sensitive and versatile.
\end{abstract}


\section{Introduction}\label{sec:intro}

The mid-infrared spectral region (5-25~\um ) is relatively underutilized for
astronomical observations, especially at high spectral resolution.
The reason for this is largely technical:
Earth's 300~K blackbody peaks in this region, resulting in high background and
high photon shot noise.
In addition, heterodyne spectrometers commonly used at longer wavelengths
are relatively insensitive in the mid-infrared, whereas dispersive
spectrographs used at visible wavelengths become quite large when used at
very high resolution in the mid-infrared.

Although difficult, mid-infrared spectroscopy provides significant, and
sometimes unique, astronomical information.
The majority of spectral features in the mid-infrared fall in three classes:
emission and absorption by grains and very large molecules; fine-structure
and recombination line emission by ions; and emission and absorption by
molecules, mostly through their vibration-rotation bands.
The spectrometers on the ISO satellite were well suited for study of
solid state features of dust grains and icy grain mantles, and the similar
features of very large molecules such as PAHs.  These features are
relatively broad ($\Delta \lambda / \lambda \approx 1-10 \% $),
so were well resolved by the Infrared Space Observatory (ISO) short
wavelength spectrometer with R = $\lambda / \delta \lambda \approx$ 2000.
SIRTF, with R $\approx$ 600 and much better sensitivity than can be
achieved from the ground, will continue this study.
However, neither of these spacecraft spectrometers was optimized for
studies of narrow gas-phase lines.

Mid-infrared ionic lines have been studied in a number of sources by several
ground-based instruments, as well as by ISO.  These lines provide information
similar to that obtained from visible wavelength forbidden and recombination
lines, but can be studied in much more obscured sources.  With widths
$\approx 10-100$ \kms\ (the latter in external galaxies), these lines are best
studied by instruments with R $\sim 10^4$.  Our previous instrument,
Irshell (Lacy et al.\ 1989), was optimized for the study of these lines.
ISO greatly expanded on the earlier ground-based work, as will SIRTF, 
although the lower resolution on the space-based mid-IR spectrometers is
insufficient for detailed kinematic studies.

Less work has been done on molecular lines in the mid-infrared.
In molecular clouds these lines have widths of only a few \kms, requiring
R $\geq$ 100,000 for optimal sensitivity and to obtain kinematic information
in line profiles.
Although a few instruments have achieved such high resolution in the
mid-infrared (heterodyne spectrometers, e.g.\ Mumma et al.\ 1982, and Fourier
transform spectrometers, e.g.\ Ridgway \& Brault 1984), the sensitivity of
these instruments was sufficient for the study of only a few of the brightest
objects in the sky.  
Considerable work has been done with ISO (e.g.\ Lahuis \& van Dishoeck 2000) 
concentrating on the many lines of molecular Q-branches.
Using Irshell, with R $\approx$ 10,000, we observed interstellar C$_2$H$_2$
and CH$_4$, as well as several other molecules, but could not resolve the 
individual lines
except in regions of shocks or rapid outflows.  As a result, we were limited
to measurements of equivalent widths of often saturated lines.
From this experience we concluded
that whereas space-based spectrometers will improve on our
observations of solid-state and ionic lines, at least in cases where high
spatial resolution is not required, a high spectral resolution and high
sensitivity ground-based (or airborne) spectrograph was required to further
the study of molecules in the infrared.

\section{Instrument Design Choices}\label{sec:choices}

The biggest difference between the mid-infrared and visible spectral regions,
from the point of view of instrument design, is the large amount of
thermal background radiation in the mid-infrared.
Consequently, mid-infrared
instruments must be cooled whenever possible.  Even with a fully cryogenic
spectrograph, photon shot noise in the background from the telescope and
atmosphere is the dominant source of noise in an instrument using a state of
the art detector.  (Heterodyne spectrometers, with additional quantum noise,
are the exception to this statement.)

Several types of spectrometers are capable of R = 100,000, with each
having advantages and disadvantages.  Heterodyne spectrometers can easily
achieve even higher resolution, but have the disadvantage of observing only
a diffraction-limited field in the sky and having quantum noise much greater
than background photon noise in the mid-infrared.
Fabry-Perot spectrometers are scanning monochromators (although with
slightly different wavelengths transmitted at different field positions),
and so must observe different wavelengths sequentially.  This inefficiency
can be overcome by the efficiency of observing many field positions
simultaneously with a detector array if information about different
field positions is of interest.  An additional drawback to the use of
Fabry-Perot spectrometers for very high resolution infrared work is that
several interferometers must be used in series to isolate a single
transmission spike of the high-order interferometer, and each of the
interferometers must be cooled to avoid excess photon noise.
Fourier transform spectrometers can also achieve the desired resolution
with sufficiently long path length modulation (1~meter for R = 100,000 at
$\lambda$ = 10~\um ).  They have the disadvantage of receiving background
radiation from the entire spectral region observed, resulting in increased
photon noise, which cancels their multiplex advantage and gives them a
sensitivity equal to that of a scanning monochromator.  
Finally, grating spectrographs also can be used for very high resolution
spectroscopy.  When used with detector arrays, grating spectrographs have
the advantage of observing many spectral elements simultaneously, while
observing a line of field positions along a narrow slit.

These sensitivity arguments led us to limit our choice of spectrometer to
Fabry-Perots and gratings. The choice between these two options was based on
the relative importance of simultaneous observations of a two-dimensional
field vs.\ simultaneous spectral and one-dimensional spatial observations.
We decided in favor of the latter, and so in favor of a grating
spectrograph, based on the fact that molecular lines, for which very high
resolution is most often needed, are often seen in absorption against
nearly point-like continuum sources.

The primary difficulty in using a grating spectrograph for very high
resolution spectroscopy in the mid-infrared is that a very large grating is
needed.  The maximum theoretical, or diffraction-limited, resolving power of a
grating is determined by the product of the number of grooves illuminated and
the diffraction order, or equivalently the optical path difference in
waves between the two ends of the grating,
\begin{equation}
\label{eqn:resol}
{\rm R}_{diff~ltd} = 2 n l / d = 2 l \sin\theta / \lambda
\end{equation}
where $n$ = diffraction order number, 
$l$ = illuminated length, $d$ = groove spacing,
and $\theta$ = incidence angle from normal.
In practice, diffraction-limited resolution is never achieved without
intolerable light loss at the entrance slit, making R $\approx l / \lambda$
a reasonable approximation.  This means that achieving R = 100,000 at
$\lambda$ = 10~\um\ requires a grating approximately 1~meter in length.

Beyond the fabrication of the grating, the primary problem in making a
very high-resolution mid-infrared spectrograph is in designing an optical
layout which uses it efficiently and which can be cryogenically
cooled in a reasonably
sized dewar.  Most high-resolution grating spectrographs use echelle gratings
with blaze angles between $\theta$ = 63.4$^{\circ}$ ($\tan\theta$ = 2 or R2) 
and $\theta$ = 76.0$^{\circ}$ (R4).  
Unless used with anamorphic optics, these gratings
and the light beams that illuminate them must have widths 1/2 - 1/4 of the
grating length for optimum illumination.  A very large dewar would be
required to contain the resulting spectrograph.  This problem can be
ameliorated by using a more steeply blazed grating.  We chose to use an R10
grating ($\theta$ = 84.3$^{\circ}$), 
requiring a light beam diameter only 1/10 the grating length.
Our grating is 36 inches long and 3.4 inches wide.
It is illuminated by a 3.3~inch diameter beam formed by a 40~inch focal length
mirror collimating f/12 light (slightly under-illuminating the echelon to
improve its efficiency at the expense of a small decrease in resolution).
The entire
instrument fits inside a 1.5~meter long, 0.4~meter diameter liquid helium dewar.

There are several drawbacks to using such a steeply blazed grating.
First, unless the grating is used in very high order the angular width of 
an order can be comparable to the angle between grazing incidence and the
blaze ($\sim 6^{\circ}$ in our case), 
resulting in a large variation in resolution across an order.
And second, the tilt of the slit image resulting from the shift in
wavelength with out-of-plane angle is proportional to the product of the
out-of-plane angle and $\tan\theta$ (the R number), and can
be very large unless the grating is used very close to Littrow.  (If instead,
input and output beams are separated in the plane of dispersion, groove
shadowing can be significant.)
We avoided the first problem by having a coarse grating with 0.3~inch 
(7.62~mm) groove spacing implying a diffraction order of 
n $\approx$ 1500 at $\lambda$ = 10~\um.
The angular width of an order, the single-slit diffraction pattern from a 
groove face, is then $\Delta \phi = (\tan\theta) \lambda / d = 0.75^{\circ}$.
This is small enough that the resolution variation across the order is
tolerable.
We minimized the second problem by using an out-of-plane angle of 0.01
radian, although the slit image rotation is still
$2 (\tan\theta) \gamma = 12^{\circ}$.

A very coarse ruling has the additional advantage of very
small order spacing.  The spectral width of a grating order is given by
$\Delta \nu = \nu / n$, or $\Delta w {\rm( cm^{-1})} = 1/2d$.  The number of
diffraction-limited resolution elements in an order is equal to the number
of illuminated grooves in the grating, or n$_{res~el} = l/d = 120$.
The small order spacing and number of resolution elements per order are
optimally matched to the 256 pixels across the available detector array,
allowing Nyquist sampling of the diffraction spot and complete coverage of an
order with appropriate optics (although not at all wavelengths since the
width of a diffraction spot and an order scale in proportion to wavelength).
Consequently, it is possible to cross disperse the spectrum and obtain
continuous spectral coverage, without gaps between orders.  
Furthermore, the echelon is held at a fixed angle, thus simplifying the 
mounting mechanism.

In the remainder of this paper we describe TEXES, the Texas Echelon Cross
Echelle Spectrograph.  TEXES uses the long, coarsely ruled grating described
here, which we refer to as an echelon, as its primary disperser and an
echelle grating used in low order as a cross disperser
to achieve our goal of R = 100,000 in the mid-infrared.  TEXES also has other
modes of operation for lower resolution observations.

\section{Design}\label{sec:design}

\subsection{Optical Design}\label{sec:optics}

TEXES is an up-looking instrument used at Cassegrain focus.  
It is in a liquid helium cooled dewar with cold fore-optics and two 
spectrograph
chambers.  The first chamber houses the high-resolution echelon grating.
The second chamber contains cross-dispersion gratings and the detector.

Figures~\ref{fig:dewar} and \ref{fig:endview} show the light path 
through the spectrograph.
Light enters the dewar through a focusable, up-looking warm lens.
This lens forms an image of the telescope entrance pupil on a cold Lyot stop.
It has litle effect on the location of the focal plane which lies just
inside the dewar.
A second (cold) lens reimages the focal plane at f/12 off the first fold
mirror, through a filter wheel, and onto the slit wheel.
ZnSe lenses are installed if the longest wavelength of interest is
$< 14$~\um, and KBr is used for longer wavelength observations.
After passing through the slit, the light enters the echelon
chamber, where it is reflected by a flat fold mirror and a 40~inch focal
length, off-axis paraboloidal collimator to the echelon grating.
The echelon is used at a steep angle, 84.3$^{\circ}$, as is discussed
in Section~\ref{sec:choices}.  
The dispersed light returns to the collimator, then to a second folding flat,
which reflects it out of the echelon chamber through a port opposite the slit.
It then enters the cross-dispersion chamber, with a similar
paraboloid-grating-paraboloid arrangement, but turned by 90$^\circ$.
The light is reflected through a focal reduction lens to the detector 
(see Figure~\ref{fig:endview}),
a $256\times256$, 30~\um\ pixel Si:As array (see
Section~\ref{sec:detector}).
The focal reducer changes the f/12 beam to f/6 so that the detector
Nyquist samples the
diffraction pattern at $\lambda \ge 10$~\um.  
A 31.6 groove/mm R4 echelle, with a length-to-width ratio of 2, 
is most often used for cross-dispersion.
The R4 grating was actually a result of a purchasing error; an R2 grating
was specified, but fortuitously R4 gratings in which both sides of the grooves
are efficient, as is the case with this grating in the infrared, have good
efficiency in Littrow configuration over a wide range of incidence angle
near R2 due to corner reflection by the grooves.
The cross-dispersion grating separates the echelon orders by $\sim 20$
diffraction-limited resolution
elements, allowing a sufficient slit length for nodding compact objects
along the slit or mapping of slightly extended objects.
The spectral coverage is typically $\lambda$/200, which includes 5-10
echelon orders, depending on wavelength.  A full echelon order lands on the
detector for $\lambda < 11$~$\mu$m giving continuous coverage
at short wavelengths, but gaps occur between orders at longer wavelengths.

Several positionable optics allow the instrument to be used in other modes,
for long-slit, lower resolution spectroscopy and for imaging.
First, the two folding flats in the echelon chamber are mounted on a
rotatable assembly.
When this assembly is rotated by 45$^{\circ}$ the incoming, diverging light
is sent to a concave spherical mirror, which
reflects it back to a convex sphere mounted on the rotating assembly
between the two folding flats.
It then returns to the concave sphere and from there to the second flat.
This arrangement forms an Offner relay, which reimages the slit onto the
output aperture of the echelon chamber, while acting like a K mirror that
rotates the slit image by 90$^{\circ}$ so that it is perpendicular to the
dispersion of the echelle grating.  In this arrangement the instrument is a
medium-resolution spectrograph with R $\approx$ 15,000 and a spectral
coverage the same as in cross-dispersed mode.
Because only one order of the echelle is used, the entire 8 mm (45\arcsec\
on a 3~m telescope) slit length can be used.
A second long-slit mode can be chosen by swinging a first-order grating into
the light beam in front of the echelle.
This 75 groove/mm R0.5 grating gives a resolution of 0.004~\um\ and a
spectral coverage of 0.25~\um\ in first order (at 8-14~\um).
The low-resolution grating can also be used as a cross-disperser for the
echelon, providing 0.25~\um\ (or $\sim$40 echelon orders) of coverage,
but requiring a very short slit or pinhole aperture to separate the orders.
In addition to the various spectrographic instrument modes, the instrument
can be used as a low efficiency camera by bypassing the echelon and
turning the low-resolution grating
face-on so that it acts as a mirror.  This mode, which is used for source
acquisition, has good image quality and 0.33\arcsec\ pixels on a 3~m
telescope, but only a
$\sim 25$\arcsec\ $\times 25$\arcsec\ field of view and 
low efficiency compared to an
optimized camera.
In a variation on the imaging mode, the K-mirror assembly can be rotated an
additional 45$^{\circ}$ for a pupil imaging mode
(see Figure \ref{fig:endview}).
A pair of flat mirrors and a lens then image the instrument
cold stop, rather than the focal plane or slit, at the output of the echelon
chamber.  With the low-resolution grating face-on,
this image of the Lyot stop and telescope pupil (the secondary
mirror) are reimaged on the detector.  This mode simplifies
alignment of the instrument with the telescope.

Raytrace calculations show the optical design is diffraction-limited
in all modes
down to 5~\um\ wavelength over a $1 \times 1$~cm ($100'' \times 100''$
on a 3~meter telescope) focal plane.
The calculated spectral resolution of the echelon when used with a
diffraction-limited slit is 0.007 cm$^{-1}$, or R = 140,000 at
$\lambda = 10\ \mu$m.
Laboratory tests with visible, near-IR, and mid-IR lasers indicate
slight phase errors over portions of the echelon.
Modeling suggests these errors increase the relative
strength of sidelobes at shorter wavelengths,
but the majority of the power should
remain in the central diffraction peak down to $\sim$5~\um\ wavelength.
Since the detector pixels Nyquist sample the diffraction pattern only
at 10~$\mu$m and longward, the resolution is effectively limited to the
10~$\mu$m diffraction limit.
For a description of the actual system performance, see 
Section \ref{sec:perform}.

\subsection{Echelon Grating}\label{sec:echelon}

The idea of making a very coarsely ruled diffraction grating was suggested
by Michelson (1898).  He attempted to make 
such a grating by stacking a number of glass
plates of equal thickness to make what was essentially a grism.
He referred to the device as an echelon, because of the staggered
arrangement of the plates.
Subsequently, reflection echelons were made (see Born \& Wolf 1980),
but were limited to a moderate number of reflecting faces.

In 1993 we undertook to assemble a reflecting echelon grating using 100
diamond-machined aluminum facets for the mirror surfaces, spaced by 0.4 in
along a 40 in aluminum support bar.  Although an echelon of this size had
not previously been made, we felt that the reduced precision required for
infrared use would make it feasible.  In fact, we demonstrated that we were
able to attach facets to the bar with the required precision.
Unfortunately, the additional requirement that the echelon had to remain
aligned during repeated cycling between 300 and 4 K for use in the thermal
infrared more than compensated for the advantage of relaxed tolerance.
The motion of the facets during cooldown ($\sim$1~\um\ rms) was several
times what we could tolerate.  The process of assembling and testing the
echelon is described in more detail by Lacy et al. (1998).

Fortunately, during the time we were attempting to assemble an echelon
grating, B. and K. Bach at Hyperfine, Inc. were developing the capability of
diamond machining gratings of this size.  In 1998 we placed an order for a 
36 inch long, 3.4 inch wide, R10 grating with 0.3 inch groove spacing.  An
excellent grating for our purposes was received in 1999.
The diamond machining and the capabilities (which unfortunately are no
longer available) of Hyperfine are described by Bach, Bach, \& Bach (2000).
(Properly speaking, the diamond-machined grating is not really an echelon,
as it is not assembled from separate mirrors,
but we continue to refer to it with that name, not wanting to have to change
the name of our instrument.)

Some care was taken in preparing the substrate before diamond machining to
avoid distortion during machining or cooling.  Our main worry was that
stress variations introduced into the substrate during heat treating would
be relieved during machining, thermal cycling, or aging.
We took several steps to avoid this.
First, the substrate was cut from a large 6061 T6 aluminum tooling
plate, oriented so that the surface to be diamond machined was inside the
plate (not on the top or bottom) so the diamond machining would not remove
material with atypical heat-treating stresses.
A 2.9-inch diameter hole was then
gun-drilled through the length of the $3.4 \times 3.4 \times 36$ inch
substrate, removing about half of its mass, but little of its stiffness.
It was then given a stress-relief anneal (650$^{\circ}$F for $\sim$2 hours,
followed by slow cooling; Gassner et al. 1964).
After that, the substrate was cycled five times between 210$^{\circ}$F and
a liquid nitrogen bath, with the heating and cooling each taking several
hours.
Finally, the grating surface was diamond machined.
In the two years since then it has been cycled between 300~K and 4~K
$\sim$15 times, with no detectable changes.

The echelon's intrinsic stiffness and the fact that it is fixed in place,
a consequence of the small angular size of the orders,
simplifies mounting.  We support the echelon purely at its ends and it
does not sag detectably under gravity, even when in a horizontal orientation
(equivalent to pointing at the horizon when on the telescope).
The mounts to the support structure are designed not to overconstrain
the echelon (see Figure~\ref{fig:dewar}).
At the top end, we use flex pivots (not shown) to constrain 4 degrees of
freedom.
At the bottom end, we use a piece of aluminum which is necked down
so that only the remaining 2 degrees of freedom are constrained.
In this piece, some effort was made to
maximize its thermal cross-section since this is the main
path for heat to flow out of the echelon.
(In retrospect, the supports should have been swapped for faster
cooldown.)

\subsection{Mechanical Design}\label{sec:mechanical}

Working in the thermal infrared near 10 $\mu m$ requires the cooling of 
instrument optics to well below liquid nitrogen (LN$_2$) temperature to
reduce background radiation.  TEXES has a
liquid helium (LHe) cooled inner chamber surrounded by walls that are heat 
sunk 
to a LN$_2$ dewar, all contained in a large vacuum jacket 
(see Figure~\ref{fig:dewar}).  
The LHe dewar bottom surface serves as
the cold surface for mounting the optical chambers.
The LN$_2$ dewar has two cold shields connected to it.
The outer shield must conduct the radiative heat load from the warm
vacuum wall along the length of the dewar up to the nitrogen vessel.
This outer shield will thus not be at LN$_2$ temperature at its
far end.
In a dewar of the length of TEXES (1.5 meter), the heat flux onto the 
LHe-cooled chamber from a single LN$_2$-cooled shield would significantly
increase
LHe boiloff.  To prevent this we 
use a second inner LN$_2$ shield attached to the bottom of the LN$_2$ 
reservoir.
Since this shield is protected from room-temperature radiation by the outer
LN$_2$ shield, it is essentially isothermal with the LN$_2$ reservoir.
To further reduce the heat load on the LHe, we pump on the LN$_2$ dewar to 
drop its temperature by $\sim 10$ K. 
The hold time of the 5~liter LN$_2$ dewar is
16-24 hours, while the 5~liter LHe dewar's hold time is about 2 days.
It takes 4-5 days, $\sim$100~liters LN$_2$, and $\sim$30~liters LHe to 
evacuate and cool the dewar to operational 
temperatures.  To minimize LHe usage, we thoroughly cold soak the
dewar with pumped LN$_2$ in the LHe reservoir, although care must be
taken to make sure no residual LN$_2$ is in the reservoir when starting
the LHe transfer.  The dewar was produced by Precision Cryogenic Systems Inc.  

G10 fiberglass supports are used to separate
the LN$_2$ dewar from the outer vacuum wall and the LHe
dewar from the LN$_2$ dewar.  The TEXES G10 support is made of 
three long tabs that connect to the vacuum jacket,
the nitrogen can twice, and then to the helium can (Figure~\ref{fig:dewar}).
This design is less prone to failure than the 6-tab used in our previous
instrument (Lacy et al.\ 1989).
However, tension in the tabs can cause them to slip at their clamps,
resulting in screw tear-out when assembling the dewar on its side, or
misalignment when on the telescope.

All internal mechanisms have external motors
driving fiberglass shafts coupled to stainless steel shafts.
Ferrofluidic vacuum feedthroughs couple the
motors with the fiberglass shafts.
There are five internal mechanisms: filter wheel, slit wheel, 
rotating K-mirror assembly, and two cross-dispersion gratings.
The motors are geared down inside the LHe temperature region
after the connection to the stainless steel
shafts, thus reducing the effect of wind-up in the fiberglass shafts.  
Both spur and worm gears are used.  
Brass wipers connected to the nitrogen dewar and pressing up against
the fiberglass shafts reduce the heat flow into the liquid helium.

All rotating mechanisms are supported by degreased ball bearings.
The K-mirror assembly and both cross-dispersion gratings use purchased ball
bearing assemblies.
The slit and
filter wheels are each supported by single ball bearings captured
along the wheel's axis.  The capture cups are machined with a center drill
in a press-fit axle for the wheel.  This design is quite simple
and takes little space.

The parabolic mirror at the end of each chamber is mounted to a backup plate
attached to the chamber end plate with a sandwich of stainless steel and
copper leaf springs.
The mirrors rest on three screws that can be adjusted by retractable
screwdrivers that enter the vacuum chamber through o-ring seals.
On the three-point mounts the mirrors can be moderately translated in
the focus direction and tipped for optical alignment.
The echelon chamber paraboloid can be tipped
to observe wavelengths that would normally fall off our detector,
which can occur at wavelengths longer than 11~\um.

The slitwheel is made of a 3.5 inch diameter brass worm-wheel,
a slit mask, and a brass insert that mates the
two together.  There are 18 slit positions on the mask spaced at
20 degree intervals: open, 3 pinholes,
7 long slits, and 7 short slits.  The slits were made using
photolithography on a 0.1 mm thick piece of brass by Photo Design of
Arizona, Inc.  
The slit mask is soldered to the backing insert, which is screwed
into the worm wheel.  Slit widths range from 120~\um\ (2 pixels) to 
550~\um\ (9 pixels).

The filter/decker wheel comes just before the slit wheel.
Filters are needed to isolate echelle orders.
There are seven discrete filters and two circular variable filters (CVFs),
all purchased from the (no longer available) stock filter selection of
Optical Coating Labs Inc.
Each discrete filter is covered with a decker mask, which defines four
different slit lengths for high-resolution mode and a long-slit position
for medium and low resolution modes.
Different slit lengths are needed in high resolution mode because of the
varying angular separation between echelon orders as the wavelength or
echelle angle is changed.
Typical lengths range from 6\arcsec\ at short wavelengths to 
12\arcsec\ at long wavelengths.  
The CVFs each have a single slit length appropriate 
for the short wavelength end of their range (4.5\arcsec\ and 6\arcsec).  
The available filters and slit lengths are listed in Table~\ref{tab:filters}.

\section{Data Acquisition System} \label{sec:daq}

\subsection{Detector} \label{sec:detector}

The detector, electronics, and software for TEXES are to be shared with
EXES, a very similar spectrograph being developed for use on SOFIA, the
Stratospheric Observatory for Infrared Astronomy (Becklin 1997).
The combined science goals for EXES and TEXES
drive both our choice of detector and the requirements for the readout
electronics.
For the detector, the requirement that EXES have good 
sensitivity at high spectral resolution in the
5.5~\um~$\leq\lambda\leq7$~\um\ range was a critical factor.
The combination of high spectral resolution, short wavelengths,
and detector response as a function of wavelength results in a low
rate of background electrons: 350~$e^-$~s$^{-1}$~pix$^{-1}$. 
To ensure our instrument is background-noise-limited, 
we must have a detector with very low dark current and
read noise.

We chose the Raytheon 256$^2$ Si:As IBC ``SIRTF'' array with the CRC-744
readout integrated circuit (Wu et al.\ 1997).
We received our science grade array in December 1999.  
According to the test results
prepared by Raytheon, the mean read noise measured 15.6 e$^-$ with
4 Fowler pairs for an 11 second integration, 
which means that the system can be background-limited,
even at high resolution at the short wavelength end of its operating range.
The quantum efficiency measured at 7.74~\um\ is almost 50\%.
Based on measurements of similar devices, the quantum efficiency should peak
near 60\% at $\lambda$ = 10-20~\um, and be above 30\% over the 5-25~\um\ range.
The pixel-to-pixel variations in quantum efficiency are $\pm$3\% (1$\sigma$).
The well size is $\approx$ 190,000 e$^-$ at 1.0 V bias.  

The relatively small well size presents a challenge for lower resolution
modes.
When operating in low-resolution and camera modes at long
wavelengths, the incident photon flux incident can saturate
the detector before we can read the entire array.
The number of electrons generated by background sky photons
[$e^-$ s$^{-1}$ pixel$^{-1}$] is given by
        \begin{equation}
        \label{eqn:epers}
N_{e^-} = {\varepsilon \lambda^2 B_\lambda(T) A \Omega \over
h c R}\eta_{opt}\eta_{det}
        \end{equation}
where $\varepsilon$ is the emissivity (instrument + telescope + sky),
$\lambda$ is the wavelength of interest,
$B_\lambda(T)$ is the Planck function evaluated at the telescope
temperature, $A$ is the telescope area, $\Omega$ is the solid angle in the
sky seen by a pixel, $h$ is Planck's constant, $c$ is the speed of light,
$R$ is the resolving power of the spectrograph, $\eta_{opt}$ is the
optical efficiency of the instrument, and $\eta_{det}$ is the quantum
efficiency of the detector.  Taking the worst case of
$\varepsilon=1$ (our blackbody calibrator;
see section~\ref{sec:calib}), $\lambda=17$~\um,
$T=275$~K, $A=71,000$~cm$^2$, $\Omega=0.33$~square arcseconds, $R=3600$,
$\eta_{opt}=0.2$, and $\eta_{det}=0.6$, we find a background count rate
of $3.6 \times 10^6 $~$e^-$~s$^{-1}$~pix$^{-1}$.  
If we can read each pixel before it accumulates 100,000~$e^-$, a 36 Hz
frame rate, then
we will safely remain in the detector's linear response regime.
The minimum settling times after switching the CRC-744 multiplexer addresses
lead to a maximum full-frame rate of around 15 Hz.  Therefore, the
low-resolution modes require partial array readouts, which
increase the frame rate almost
inversely to the fraction of rows actually read out.

The CRC-744 chip carrier mounts in a 3M Textool socket, the lid of which has
been opened at the center to allow light to hit the detector and relieved near
the edges to avoid short circuits on the front-surface contact pads. 
The socket mounts to a G10 header board, which is attached to the
LHe-cooled aluminum wall of the cross-dispersion chamber, with the copper
ground plane of the board in contact with the wall.
When the detector is powered, the board comes to a temperature about 5~K.
We do not have a temperature sensor on the detector chip, but from the
detector time constants there is evidence that it is running at a
temperature somewhat below the optimal 6-8~K.

\subsection{Electronics} \label{sec:electronics}

To run our detector, we purchased a version of the IR Observer$^{\rm TM}$ 
electronics system from Wallace Instruments.
The system includes an analog box with clock driver boards, DC bias boards,
and  preamp boards; a digital box with timing-pattern control boards,
14-bit analog-to-digital converters, and a data storage
and coadder unit; a
rack-mounted control computer running Linux; a rack-mounted power supply;
and all the cables required to interface the system with the cryostat.
The system came equipped with software suitable for lab testing; we
describe some of the modifications required for use at the telescope below
and in Section~\ref{sec:software}.  

The electronics step through and read out the detector pixels continuously.
The signal from each pixel
is amplified in the analog box and transferred to the digital box where
it is converted to 14-bit digital data and stored as a 32 bit integer
in the coadder buffer.
The number of frames combined in the coadder, ``hardware coadds,''
is set by the user.  Once the appropriate number is reached, the coadder
begins filling a second buffer while the first is asynchronously transferred
to the computer.  Collecting hardware coadds currently establishes the
minimum unit of time for acquiring data.  When nodding the telescope, the
continuous detector readout means at least two frames
must be discarded to ensure valid data on every pixel.
We added a level of ``software coadds'' in the computer to shorten the
minimum time unit and improve our observing efficiency.  The computer writes
the data to disk after the desired number of software coadds.

Both the array clocking patterns and the patterns for clocking data into
the coadder are controlled by a set of EPROM programs resident in the
digital box.
Two levels of EPROM programs contain the basic clocking information.
The ``Master'' EPROM, a single 8 bit by
256kB chip, contains a series of starting addresses for separate slave 
routines.
The five 8 bit by 256kB ``Slave'' EPROMs operate in parallel and control the
clocking levels, the coadder synchronization, and can control a 
chopping secondary. 
The hardware can contain 32 8kB master patterns that can select from
32 8kB slave patterns.  The patterns
currently programmed into the EPROMs include a double sampling 
(DS) mode, a simple read and reset (single sampling) mode, and
partial array readouts.
The CRC-744 resets entire rows at a time so that the
DS data taking is achieved by reading an entire row, resetting it, and
reading it again.
In addition, we are testing a Fowler pattern with 
16 pairs.
If more than 32 patterns are needed, additional EPROM chips must be
programmed and swapped into the control board. 

\subsection{Data Acquisition and Display Software} \label{sec:software}

The data acquisition software for TEXES is divided into two regimes:
data taking, involving instrument control and array readout, and data display.
Each regime is controlled by a separate computer and is written in
a separate software language, determined by the needs of the system.
Communications between the two regimes are minimized, allowing for
separate development and refinement of each branch of the software.
Since TEXES will always be run by someone familiar with the project,
our primary goal is for the software to be efficient, with ease of
learning being a secondary consideration.  We expect to
continually improve the software over the life of the instrument.

Data acquisition is governed by the Linux computer mounted on the telescope.  
This computer controls the detector, operates the instrument, and
interacts with the telescope control system.  The software provided by 
Wallace Instruments 
for running the detector is written in \tcltk\ and \C.  We chose to
write the other control routines in these languages as well.  
For all control routines, our goal is to allow input from the command 
line, a graphical user interface (GUI), and scripts.

Motors can be driven locally from a front panel on
the electronics rack or remotely by the computer.
The ``home-grown'' control circuit multiplexes three binary input lines to
the six motors with addresses for stop and reset.  The control circuit
passes pulses to individual Vexta motor controllers.  When operating
remotely, a National Instruments analog and digital I/O board provides
the pulses.
Registration of the mechanisms is done by counting motor steps and
reading, with the same A/D board, 10~volt potentiometers linked to the
mechanism gear trains.  The user is alerted if there is a discrepancy between
desired and actual voltage.

The instrument control computer communicates with the telescope to command
beam switches and obtain header parameters.
At the IRTF, we establish a pipe to use the existing telescope control 
program.

Once the data have been saved to disk by the instrument control computer,
they also are passed to the data reduction program.  Our quicklook data
reduction is written in IDL running on a Sun workstation
and can be operated from the command line or from a GUI 
(Figure~\ref{fig:gui}).  
The GUI has three main display windows: the image of the detector, 
a collapse of the image along the vertical dimension (the spectral window), 
and a collapse along the horizontal direction (the spatial window).  
The quicklook roughly corrects the display so that
dispersion runs nearly horizontally and the slit runs nearly vertically.  
The user can interactively select portions of the detector image to show
in more detail in the spectral and spatial windows.  
It is also possible to display a preliminary spectral extraction of one order.
At any time while the data are being collected, the user can switch between
displaying a single raw frame, the difference image of a nod pair, or the
sum of all nod pair difference images in the current observation.   

\section{Observations and Data Reduction} \label{sec:obs}

\subsection{Observing Modes}

TEXES has four basic observing modes: stare, nod, map, and scan,
each of which can be used with any of the spectroscopic or imaging
instrument modes.

Stare mode involves acquisition of individual frames or coadded sets of
frames without any telescope motion.  It is used mostly for diagnostic
purposes and not for astronomical observations.

In nod mode, the telescope is moved between coadded sets of frames to
alternate the object either between two positions along the slit (or within
the imaging field) or on and off of the slit.
Typically, 4-6 seconds are spent in each nod position, and one second is
spent moving the telescope and waiting for it to settle after each motion.
Unlike mid-infrared imaging systems for ground-based telescopes, we do not use 
the chopping secondary.
In most weather conditions the systematic fluctuations in the sky emission
(sky noise) are less than the photon shot noise at the $\sim$1/12~Hz nod
frequency.  Even in non-photometric weather, sky noise is largely removed by
differencing of the two nod beam measurements when nodding a source between
two points along the slit.
We have experimented with multiple-beam nod patterns, but do not currently
allow this option.

In map mode, the telescope is stepped between nod pairs.
The nod throw takes the slit entirely off of the object,
and the sky frames preceding and following each source frame are averaged
to reduce their contribution to the noise.
This mode is used to map extended objects or to
search for bright objects.

Scan mode is a variation on map mode, but without telescope nodding.
The telescope is stepped a long enough distance so that blank sky is
observed at one or both ends of the scan, and those observations are used as 
sky
measurements and subtracted from the on-source measurements.
In clear weather scan mode can be very efficient.  The time spent on
sky can be small compared to the total time spent on the source being mapped,
but large compared to the time spent on each position of the map.
Therefore, the
noise in the average of the sky frames is small compared to the noise in the
source frames from which it is subtracted.

\subsection{Calibration} \label{sec:calib}

The presence of strong background emission from the sky and telescope limit 
sensitivity in the mid-infrared.  This disadvantage is partially offset 
by the radiometric calibration that this emission makes possible.  We 
calibrate the measured source emission in a manner very similar to the 
radiometric scheme used at millimeter and submillimeter wavelengths (e.g.
Ulich and Haas 1976).
At the beginning of each observing sequence (typically every 10 minutes)
a set of calibration frames is taken.
These consist of measurements of a black chopper blade, the sky, and a
low-emissivity chopper blade just above the dewar entrance window.
The standard procedure is to use the $black - sky$ difference as a flat
field.  To the approximation that the temperatures T$_{black}$,
T$_{telescope}$, and T$_{sky}$ are equal, this flat field serves to
correct both for spectral and spatial instrumental gain variations and for
the absorption (or blockage) of light by the telescope and the atmosphere,
as can be seen from the following:
\begin{eqnarray}
\label{eqn:flat}
{\rm S}_{\nu} (black-sky) & = &
\{ {\rm B}_{\nu} ( {\rm T}_{black} )
- [ {\rm B}_{\nu} ( {\rm T}_{tel} ) \varepsilon_{tel}
+ {\rm B}_{\nu} ( {\rm T}_{sky} ) \varepsilon_{\nu,sky}
( 1 - \varepsilon_{tel} ) ] \} {\rm R}_{\nu} \\
& \approx & {\rm B}_{\nu} ( {\rm T}_{tel} )
( 1 - \varepsilon_{tel} ) ( 1 - \varepsilon_{\nu,sky} )  {\rm R}_{\nu} \\
& = & {\rm B}_{\nu} ( {\rm T}_{tel} ) {\rm t}_{tel} {\rm t}_{\nu,sky}
{\rm R}_{\nu},
\end{eqnarray}
and consequently the calibrated source intensity corrected for
the transmission of the telescope and sky is given by
\begin{equation}
{\rm I}_{\nu} (obj)
\approx {\rm S}_{\nu} (obj-sky)
\times \frac { {\rm B}_{\nu} ( {\rm T}_{tel} ) }
{ {\rm S}_{\nu} ( black-sky ) },
\end{equation}
where ${\rm S}_{\nu}$ is the measured signal,
${\rm B}_{\nu} ({\rm T})$ is the black-body function at temperature T,
$\varepsilon$ is the emissivity of the telescope or sky,
R${_\nu}$ is the instrumental responsivity, and
t = 1 - $\varepsilon$ is the transmission of the telescope or sky.

The flat-fielding procedure reduces the depths of H$_2$O lines by
$\sim$80\%, CO$_2$ and N$_2$O lines by $\sim$50\%, and O$_3$ lines by
$\sim$20\% (since these molecules are low, mixed, and high in the
atmosphere, respectively, and so are at different temperatures).
It should be possible to remove the residual atmospheric absorption by
dividing by comparison stars.
Unfortunately, we have found that nearly all stars bright enough to serve as
good comparisons (mostly K and M giants and supergiants) have photospheric
or circumstellar H$_2$O, with lines throughout the mid-infrared.
Instead we have begun to use asteroids as atmospheric comparison sources.
They appear to work well for this purpose, especially at our longer
wavelengths where they are bright.

\subsection{Pipeline Data Reduction}\label{sec:pipe}

The data stored during observations are from the Wallace Instruments
coadder, with both sides of each nod pair appended to a disk file during an
observing sequence.
A pseudo-FITS header is stored separately from each binary data file.

First-pass data reduction is carried out by a pipeline reduction program
written in Fortran.
The pipeline reduction involves the following steps:
\newline
1) read in the calibration (flat field) frames and their header,
\newline
2) use the $black$ frame to locate echelon orders and check distortion
parameters, particularly the order rotation caused by the K mirror,
\newline
3) derive a calibration image from $black - sky$,
\newline
4) read in the object header and data frames,
\newline
5) search for and correct spikes in individual frames,
\newline
6) optionally attempt to correct for optics bounce,
\footnote{
In cross-dispersed mode, very small shifts in the optics between nod beams
can cause spurious signals from imperfect background cancellation
in the regions of steep variation of the
background at the edges of echelon orders (see Section~\ref{sec:perform}).
We have had some success in removing this effect by shifting or smoothing
the measured data in one nod beam relative to the other before differencing.
However, a nodded source can be interpreted by the software as optics
bounce, especially if it comes near the ends of the slit, so bounce
correction must be used with caution.}
\newline
7) difference the two beam measurements of each nod pair,
\newline
8) interpolate over dead or noisy pixels,
\newline
9) calibrate and flat-field the data frames with the calibration frame,
\newline
10) correct for optical distortions (this is based on data stored in the
headers and knowledge of the optical design of the instrument, and is
accurate to better than a pixel over most of the array),
\newline
11) optionally shift each nod difference frame in the spatial direction to
correct for guiding errors along the slit,
\newline
12) add nod pairs together, optionally weighted by the continuum signal in
each nod pair,
\newline
13) extract a spectrum from the difference frame, optionally weighted by the
distribution of continuum signal along the slit,
\newline
14) optionally ask the user to mark the position of an atmospheric line in
the calibration frame to set the wavenumber scale,
\newline
15) store the difference frame and extracted spectrum, and
\newline
16) optionally add the difference frame and extracted spectrum to a running
average for an object.

An important feature of the pipeline reduction is that it requires very
little user intervention, while accurately correcting for optical
distortions, deriving a linearized spectrum, and optimally weighting nod
pairs and the distribution of signal along the slit to maximize the final
signal-to-noise ratio.
Normally the user simply makes up a script file with flags for options and
names of files to be reduced, then runs the program.
User intervention is required only to specify an atmospheric line for
accurate wavenumber calibration.
The pipeline reduction requires about one hour of Sun ultra 5 CPU time to
reduce a night's data.
Data can be reduced during or immediately after
a night of observations, so instrument users will leave the telescope with
understandable spectra.
Further custom reduction, $e.g.$, selection and combination of
observation sets, customized extraction of spectra, and division by
comparison objects is done with whatever software the user chooses.

\section{Performance} \label{sec:perform}

After our first experience at the telescope, we found that astronomical
objects generally are poor tests for TEXES's spectral resolution. 
Therefore, we made a gas cell that mounts over the dewar entrance lens
to test the spectral resolution, internal instrument focus, and
mechanism zero-points.
The cell consists of a teflon cylinder of $\sim$2~inch diameter and 1~inch
length, which is sealed to the window with an o-ring, and which is
terminated with a mirror that looks back into the cold dewar.
The cell can be evacuated and then filled with low pressure gas.
The mirror arrangement provides a low background, allowing very weak molecular
thermal emission lines to be seen much more easily than in an absorption
cell.
Example emission-line spectra of CH$_4$ (7.67~$\mu$m) and 
C$_2$H$_2$ (13.7~$\mu$m) are shown in Figure~\ref{fig:lab-spec}.
Gaussian fits to the data give
a FWHM of 0.007~cm$^{-1}$ at 13.7~$\mu$m (R=100,000).
Shortward of 8~\um, the best fit Gaussian has a width corresponding 
to R=75,000, but the line profile systematically has extra power in a 
red wing.  This behavior is expected
given the groove-spacing errors in the echelon.

The response of the instrument, both to blackbody radiation and to
starlight, is consistent with an overall efficiency of the optics and
detector of $\sim$10\% at wavelengths near 8 and longward of 12~\um.
However, the efficiency drops by a factor of up to 4 in a broad trough
between these wavelengths (Figure~\ref{fig:rqe}), an effect 
we do not yet understand, but are investigating.  
In high-resolution, cross-dispersed mode the measured noise is consistent
with the calculated photon shot noise, given the measured efficiency.
However, much higher noise is observed in medium- and low-resolution modes.
The output of the 
detector array becomes very spiky above a photo-current threshold near
the level observed in medium-resolution mode, increasing the noise by more
than an order of magnitude.  Reducing the effective bias across the
array improves the behavior, but compromises the response in high-resolution
mode.  Until we find a better solution, we plan to adjust the bias depending
on the observing mode.

Figure~\ref{fig:nefd} shows the Noise Equivalent Flux Density (NEFD)
for our high-resolution mode.  These data are derived from signal-to-noise
measurements when observing
a blackbody and calculating the NEFD taking parameters appropriate to
TEXES on the IRTF: system emissivity of 0.10, 
atmospheric transmission of 100\% (true in much of the window), 
slit losses of 20\%, resolving power of 100,000, and a 3m telescope.
The NEFD is appropriate for on-source time while nodding the
telescope.  A nodding efficiency
parameter (30--80\%\ depending on conditions and target extent)
must be included when estimating clock time. 
The NEFD will scale 
roughly as the square root of the total emissivity (including the
increased emissivity from the atmosphere on terrestrial lines)
and inversely with the telescope diameter squared.  Therefore, TEXES 
observing on an 8-10m telescope with 5\%\ emissivity 
(our foreoptics contribute 5\%\ emissivity) should give an order 
of magnitude improvement in the NEFD or a factor of 100 in time.

We have conducted numerous tests on repeatability of mechanisms
and the dependence of flexure of elements with time and telescope position.
The slitwheel motions are reproducible to the one pixel level.
Filter positioning is
satisfactory with respect to the choice of slit length, but positioning
of the CVF to maximize throughput still must be done by finding
half power points.  This may simply be the result of an incorrect
dispersion relation in the CVF positioning software, but further
tests need to be done.  The echelle and first-order grating come to 
the desired position within a few pixels.  We typically confirm grating 
positioning using sky frames that show telluric emission lines.  
The K-mirror assembly is the most prone to positioning errors,
primarily due to backlash in the gear train.  Although we have generally
tried to remove backlash by forcing the motor control software to
always travel the same direction before stopping, something
more must be done for this mechanism.

In general, flexure within the instrument and between the instrument
and the telescope guide camera is acceptable.  By imaging a pinhole
and slewing $\pm$1.5~hours East-West from zenith, we determined the 
internal flexure between
our slit and the focal plane is $\approx$1~pixel for the full 45$^{\circ}$
motion.  After peaking on the spectral signature of a bright star and
setting the telescope camera crosshairs on the visible signal, a 60$^{\circ}$
motion (EW) resulted in a 1\arcsec\ error in positioning while a 40$^{\circ}$
motion (NS) resulted in a 2\arcsec\ error.  During this test, the
seeing was fairly poor, $\approx$1.3\arcsec, so while the results are 
not definitive, they are certainly tolerable.

The major problem with flexure, or possibly other internal motion, comes
with the echelle grating or its associated paraboloid.  When observing
the same source for 1~hour, a 1~pixel shift is apparent along the
echelle dispersion direction.  Because our flatfielding and atmospheric
corrections are done with calibration images taken at the start of
each observation (Section~\ref{sec:calib}), this drift effectively
limits observations to $\sim$15~minutes before additional calibration
images are required.  We see no such drift in the direction of the
echelon dispersion, suggesting the echelle grating is moving.  

A more serious problem can arise when the echelle grating appears to
bounce in response to telescope nodding.  
Because the contrast between illuminated and 
unilluminated regions of the focal plane is so large, any 
shifts in the illumination pattern between the two nod beams
results in large systematic features in the difference image.
We have seen two types of bounce: one consistent with a systematic
shift in the echelle's position between the two nod beams and the
other consistent with a shaking of the echelle throughout one of the
two nod beams.  
These problems seem to be more common
when observing near the meridian at substantial airmass and nodding
along a north-south slit.
We have not had the engineering time required to test the conditions for
causing the bounce, but have developed a software fix
(see Section~\ref{sec:pipe}).

\subsection{Selected Astronomical Observations } \label{subsec:data}

To date, TEXES has had 8 nights on the McDonald Observatory 2.7~m and
35 nights on the NASA IRTF for a mix of engineering and science.  
We present here observational results from some of the projects 
in progress.  

In Figure~\ref{fig:irc2}, we present data from the project for which
the instrument was designed: molecular absorption along the line-of-sight
to embedded star forming regions.  Two orders out of the six observed
toward the KL region of Orion
are presented; one with the C$_2$H$_2$ R(7) line and one with the HCN
R(11) line.  Both lines are clearly resolved and show non-Gaussian
line profiles composed of several very narrow components.  
These facts contradict assumptions in 
analysis of Irshell data (Evans, Lacy, \& Carr 1991; Carr et al. 1995).  
These data came from less than
20~minutes of integration time.  We have substantially more time in November
2001 at the IRTF to examine the absorption behavior toward Orion
in more detail.  Other targets will be presented in a paper by Evans et al.
(in prep).

As mentioned above, most late-type stars are full of spectral
features in the mid-infrared; Figure~\ref{fig:aboo} illustrates this point
with a ``hi-lo'' (high-resolution grating cross-dispersed by
the first order, low-resolution grating) spectrum of Arcturus.  
Arcturus is a K1.5III star and so is moderately hot by the standards of
mid-infrared spectroscopy.  
We show 20 of the
22 orders recorded in a single grating setting with terrestrial lines
marked and many stellar lines labeled.  
The strongest emission lines are from Mg~I and the strongest
absorption lines are OH.  Most of the unlabeled absorption lines
come from H$_2$O.  This observation contains 180 seconds on-source 
integration time.  The emission lines may be important to magnetic
field measurements (Carlsson et al. 1992; Valenti et al. in prep).
Lacy et al. (in prep) will examine
H$_2$O abundances and kinematics for late type stars.

TEXES will be a very effective instrument for studying the atmospheres
of planets.  In Figure~\ref{fig:jupscan} we show a map of Jupiter's
northern limb summed 
over the spectral region 818.0 to 820.0~cm$^{-1}$ (12.195 to 
12.225~\um), which includes continuum and ethane line emission.  
By using the scan mode, we are able to 
observe a slice across the disk of Jupiter and look for latitudinal
and longitudinal variations.  
By observing multiple lines of various molecules	
at high spatial and spectral resolution we can investigate
the atmospheres in more detail than was previously possible (Greathouse et
al. in prep).

Another class of objects that will benefit for high spectral and spatial
resolution study in the mid-infrared is that of ultra-compact H~II regions.
These objects contain a massive star deeply embedded in gas and dust
and creating a very compact, high density region of ionized gas.  
Given the extinction, most of our knowledge of these regions comes from
radio continuum and recombination lines observations, as well as 
infrared imaging.  We observed
one such region, Mon~R2~IRS1, in the [Ne~II] fine structure line at 
780.4~cm$^{-1}$ (12.814~\um).  The observations were done using the map mode
in roughly 8~minutes of clock time.  
The velocity-integrated map (Figure~\ref{fig:monr2map}) shows structure very
similar to what is seen in the radio continuum 
and near infrared recombination lines (Massi, Felli, \& Simon 1985;
Howard, Pipher, \& Forrest 1994).  
However, the position-velocity diagram (Figure~\ref{fig:monr2pv})
demonstrates
that [Ne~II] observations give considerable additional information.
In this case, we see the line is broad and centrally peaked where the
emission is strongest, but becomes double-peaked with comparatively
little emission at the central velocity away from the peak.
Interpretation of these data will be in Zhu et al. (in prep).

\acknowledgments

The fabrication of TEXES was supported primarily by the National Science
Foundation, with additional support from the National Aeronautics and
Space Administration through the Universities Space Research Association.
Observations with TEXES were supported by the Texas Advanced Research Program.
We thank Bianca Basso, Wanglong Yu, Jimmy Welborn, George Barczak, and
Doug Edmonston for their able assistance.

\clearpage

\begin{deluxetable}{cl}
\tablecaption{Filters and Slit Lengths. \label{tab:filters}}
\tablewidth{0pt}
\tablehead{
\colhead{Wavelength Range} & 
\colhead{Slit Lengths\tablenotemark{a}}  \\
\colhead{ [ $\mu$m ] } & \colhead{ [ Arcseconds ] }
}
\startdata
7.5 - 8.5  &  4.4, 6.0, 7.7, 9.3, 45 \\
8.4 - 9.2  &  4.4, 6.0, 7.7, 9.3, 45 \\
9.4 - 11.4 &  5.0, 6.6, 8.5, 10.4, 45 \\
11.2 - 12.2 & 5.5, 7.1, 12.1, 16.5, 45 \\
11.7 - 14.2 & 6.0, 8.0, 9.9, 11.8, 45 \\
17 - 22  &    11.5, 14.6, 18.1, 22.0, 45 \\
8 - 13  &    45 \\

4-8 CVF\tablenotemark{b}  &    4.4  \\
8-14 CVF\tablenotemark{b} &    5.5  \\

\enddata


\tablenotetext{a}{Medium- and low-resolution modes use the 45\arcsec\ 
slits.  In high-resolution mode, the slit length used depends on the
angle, and so the resolving power, of the cross-disperser.  }

\tablenotetext{b}{CVFs are $\sim$1.5\%\ bandwidth.}

\end{deluxetable}

\clearpage

\figcaption[fg1.eps]{Two views of the TEXES dewar and optics, 
with many structural details omitted for clarity.
The dot-dashed line shows the path of the light through the system
in the high-resolution cross-dispersed mode.
The low-resolution grating is folded down and out of the light path in this
mode. \label{fig:dewar} }

\figcaption[fg2.eps]{View of the TEXES optics looking toward the LHe
work surface.
The two off-axis paraboloids have been omitted from this view.
The K-mirror assembly is shown rotated to the position that causes the light
to pass through the pupil imaging mirrors and lens. \label{fig:endview}}

\figcaption[fg3.ps]{The Quicklook GUI with a 
cross-dispersed spectrum of a star after differencing the two nod beams.
The image is rotated in the software to make the dispersion run nearly
horizontal.  The spectral window below the grey-scale displays an
extraction of the 6$^{th}$ order from the top.
The spatial window to the right shows the wavelength-integrated intensity as
a function of position along the slit for each order. \label{fig:gui} }

\figcaption[fg4.ps]{Laboratory spectra of the CH$_4$ Q branch near 
7.7~$\mu$m (top panels) and the C$_2$H$_2$ Q branch near 13.7~$\mu$m (bottom
panels).  The left panels show roughly 60\%\ of the coverage, and the 
right panels show Gaussian fits to selected lines for each molecule.  
The short wavelength line profiles are non-Gaussian, particularly
on the red edge.  For CH$_4$ the FWHM of the best fit Gaussian gives
R=75,000, while for C$_2$H$_2$ the FWHM gives R=100,000.
\label{fig:lab-spec} }

\figcaption[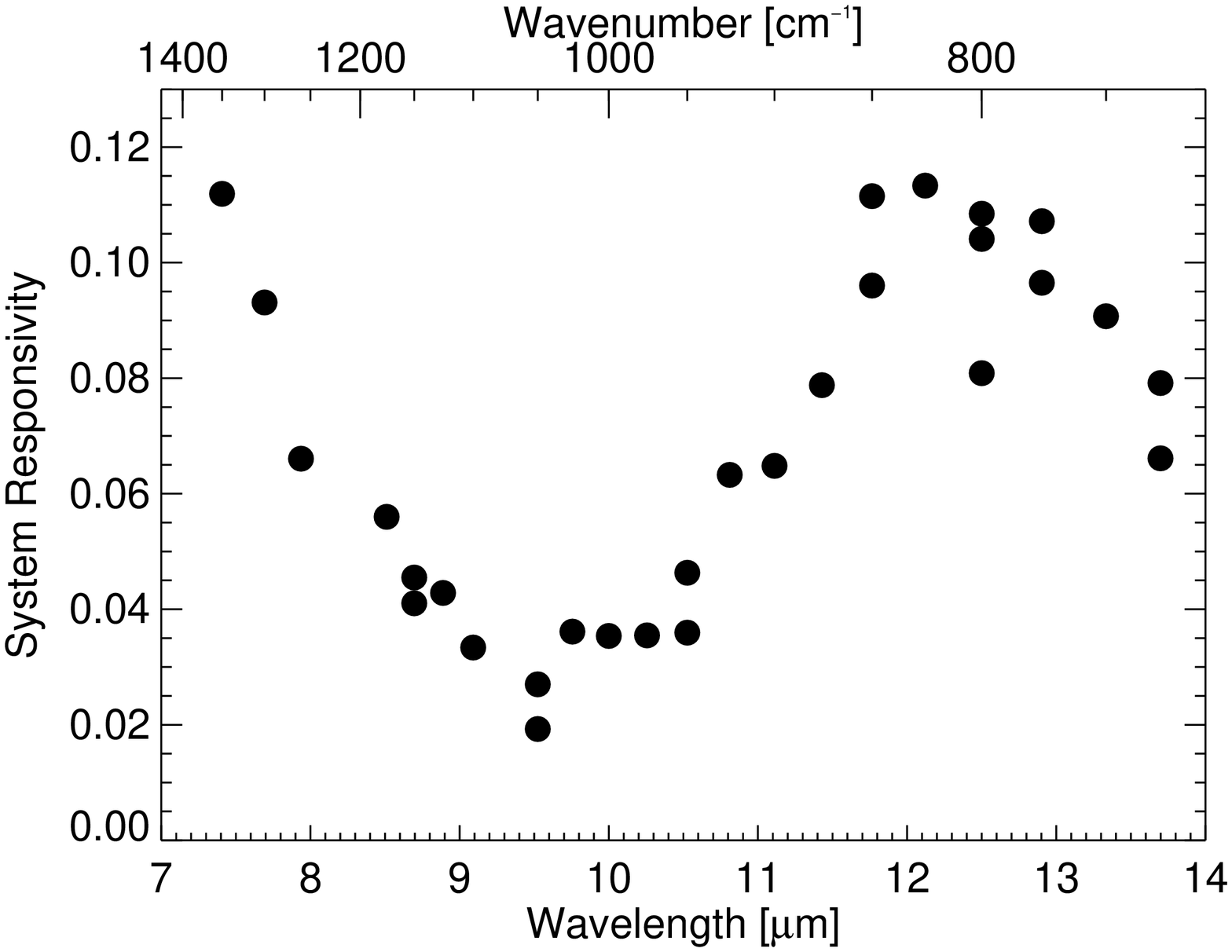] {TEXES's system response in the 8-13~$\mu$m range
as determined from a blackbody.  This response is near the peak
of the echelon blaze.  For wavelengths far off blaze, the response can
be a factor of 2 lower.\label{fig:rqe} }

\figcaption[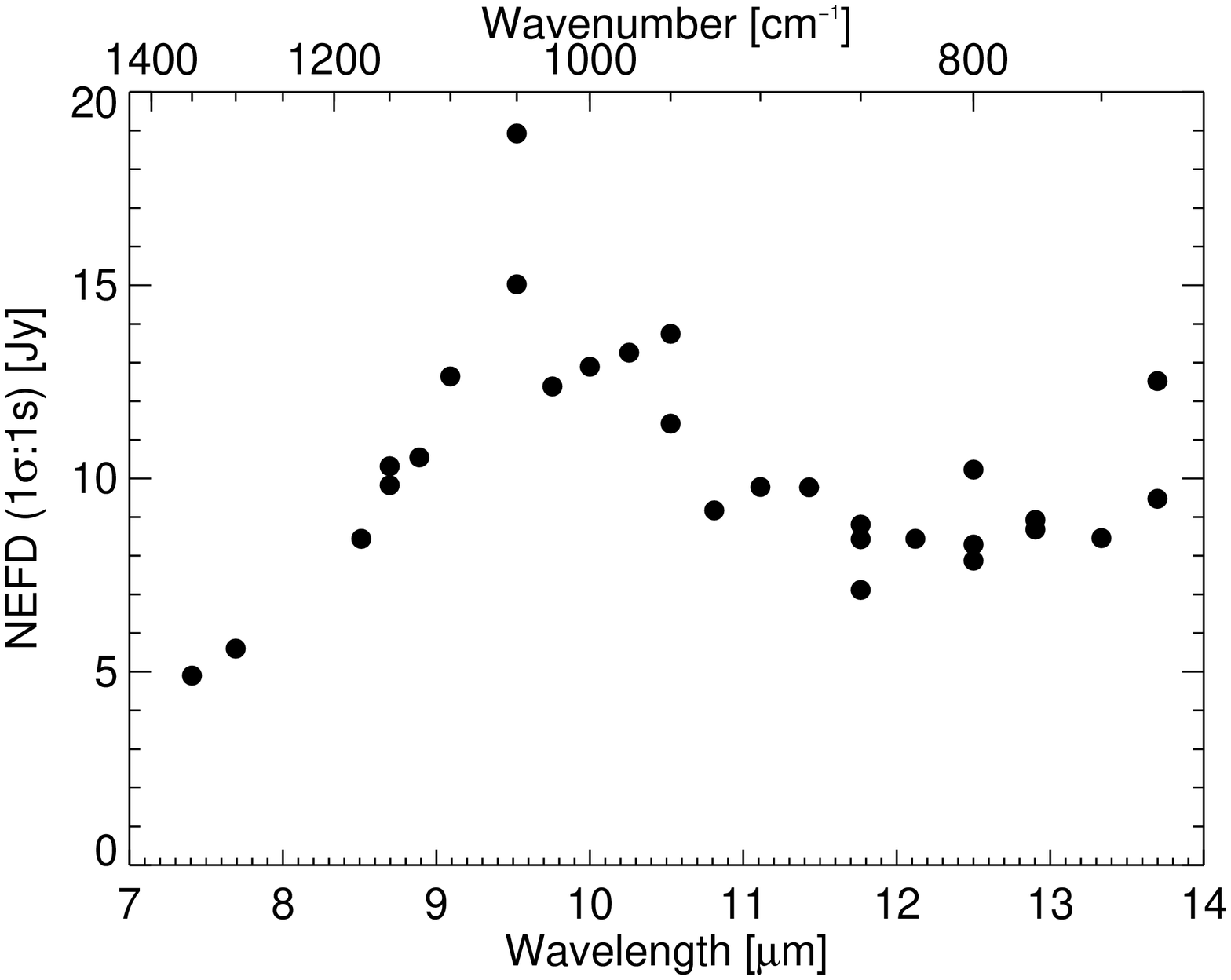] {Calculated NEFD based on 
blackbody measurements and assumptions as given in the text.\label{fig:nefd} }

\figcaption[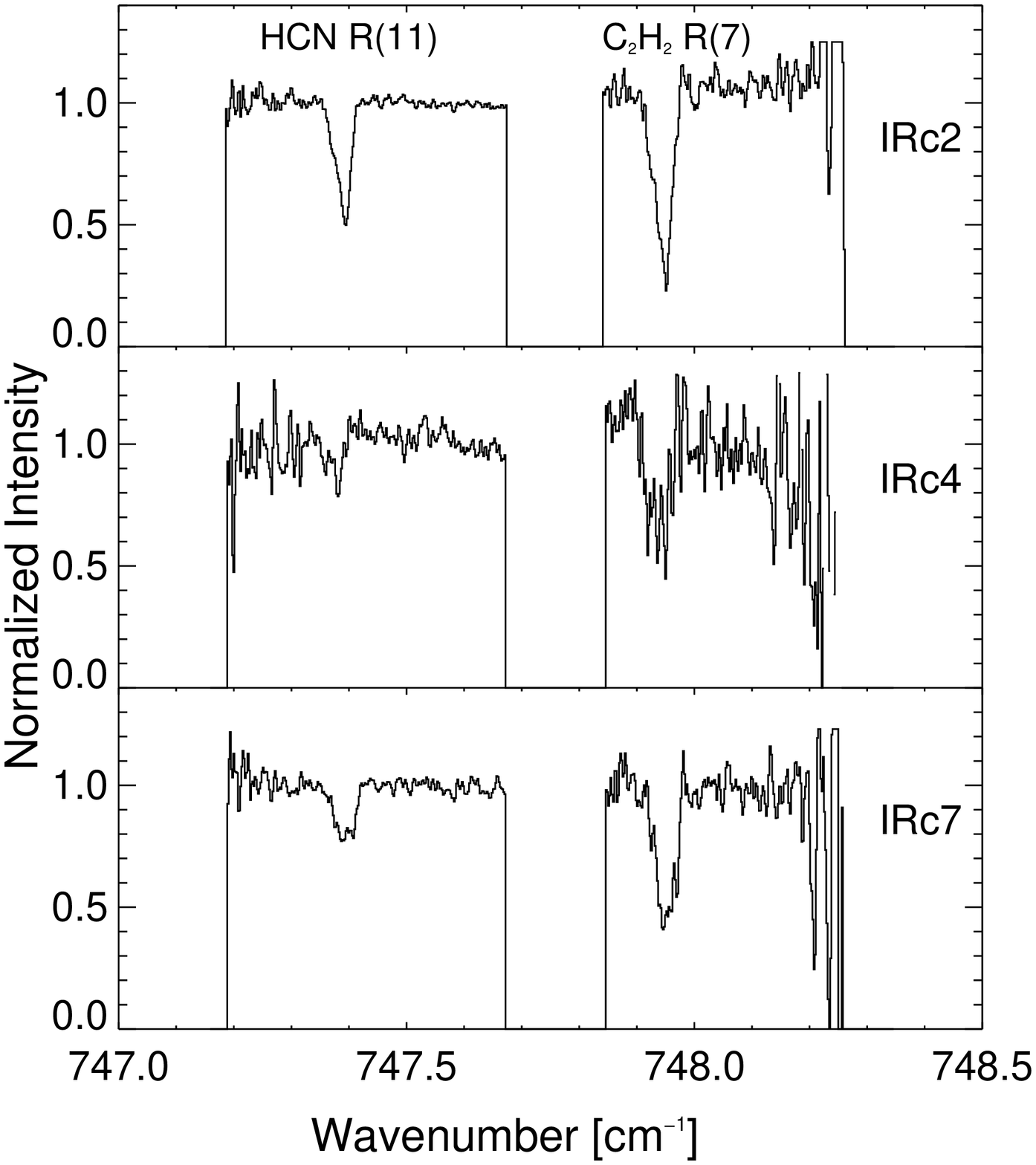]{High-resolution spectra showing
C$_2$H$_2$ and HCN absorption along the line of
sight to Orion IRc2, IRc4, and IRc7.
Two orders out of six are displayed in each case.
Atmospheric emission and absorption cause the rise in noise near the ends of
the region plotted, but the structure near the C$_2$H$_2$ line is real.
\label{fig:irc2} }

\figcaption[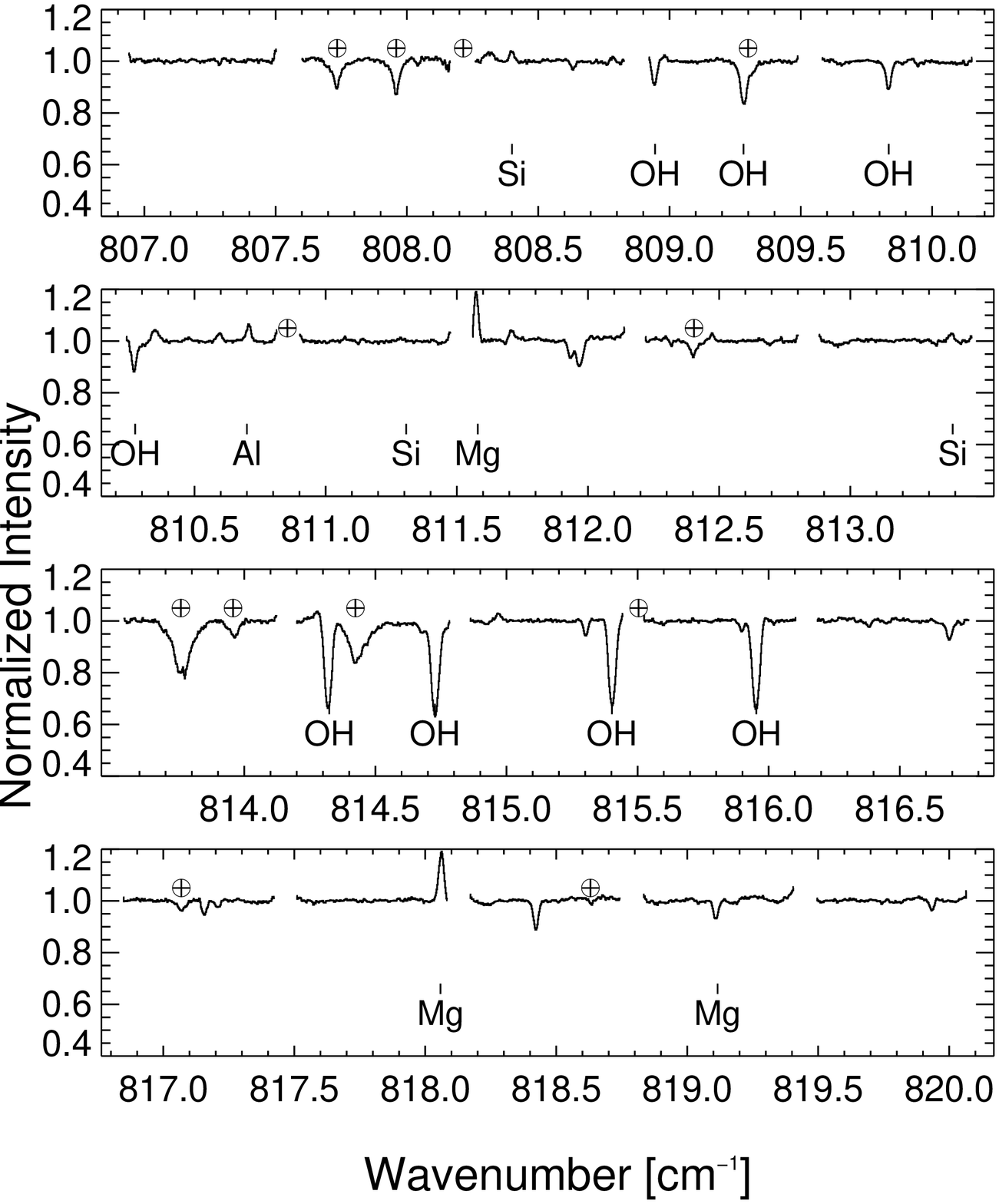]{High-resolution spectrum of Arcturus using the
first-order grating as the cross-disperser.
Telluric lines are indicated ($\oplus$).
Some stellar features are labeled, with  
most of the unlabeled absorption lines being due to water.
\label{fig:aboo} }

\figcaption[fg9.ps]{Top panel:  Scan across the northern limb of Jupiter.  
The slit was oriented along the planetocentric N/S direction and was stepped
${2 \over 3} ^{\prime\prime}$ from east to west.  The bright
region is an enhancement due to the North Auroral Ring.
Below: Ethane spectrum corresponding to a pixel in the 
middle of the bright enhancement above.    
\label{fig:jupscan} }

\figcaption[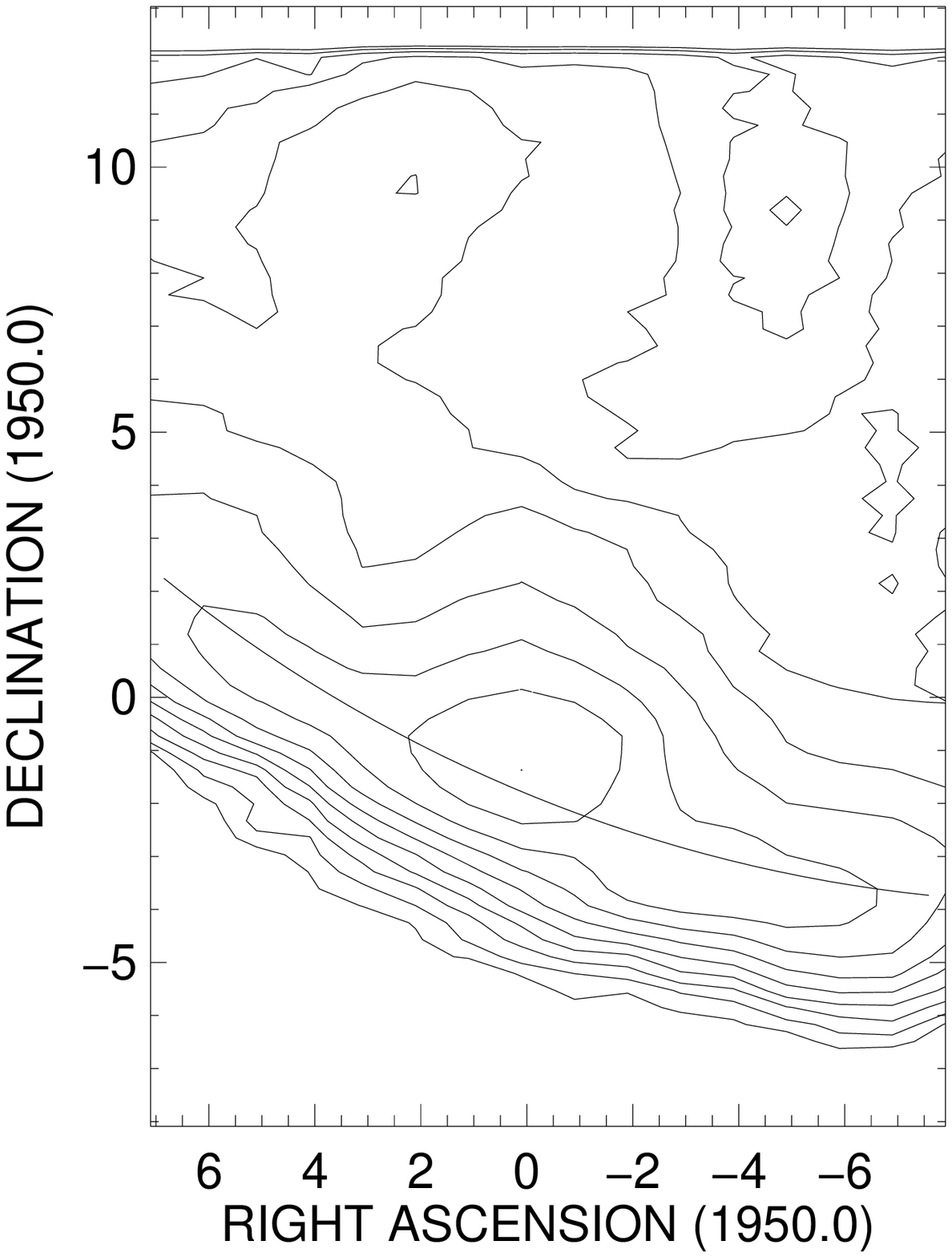]{[Ne~II] (12.8$\mu$m) line map of Mon R2 IRS1
integrated over all velocities.  
Contours are drawn at 0.5, 0.35,
0.25, 0.17, 0.125, 0.088, 0.0625 and 0.044 of the peak value.
Pixel values along the curve are used for the position-velocity
diagram in Figure \ref{fig:monr2pv}.
\label{fig:monr2map} }

\figcaption[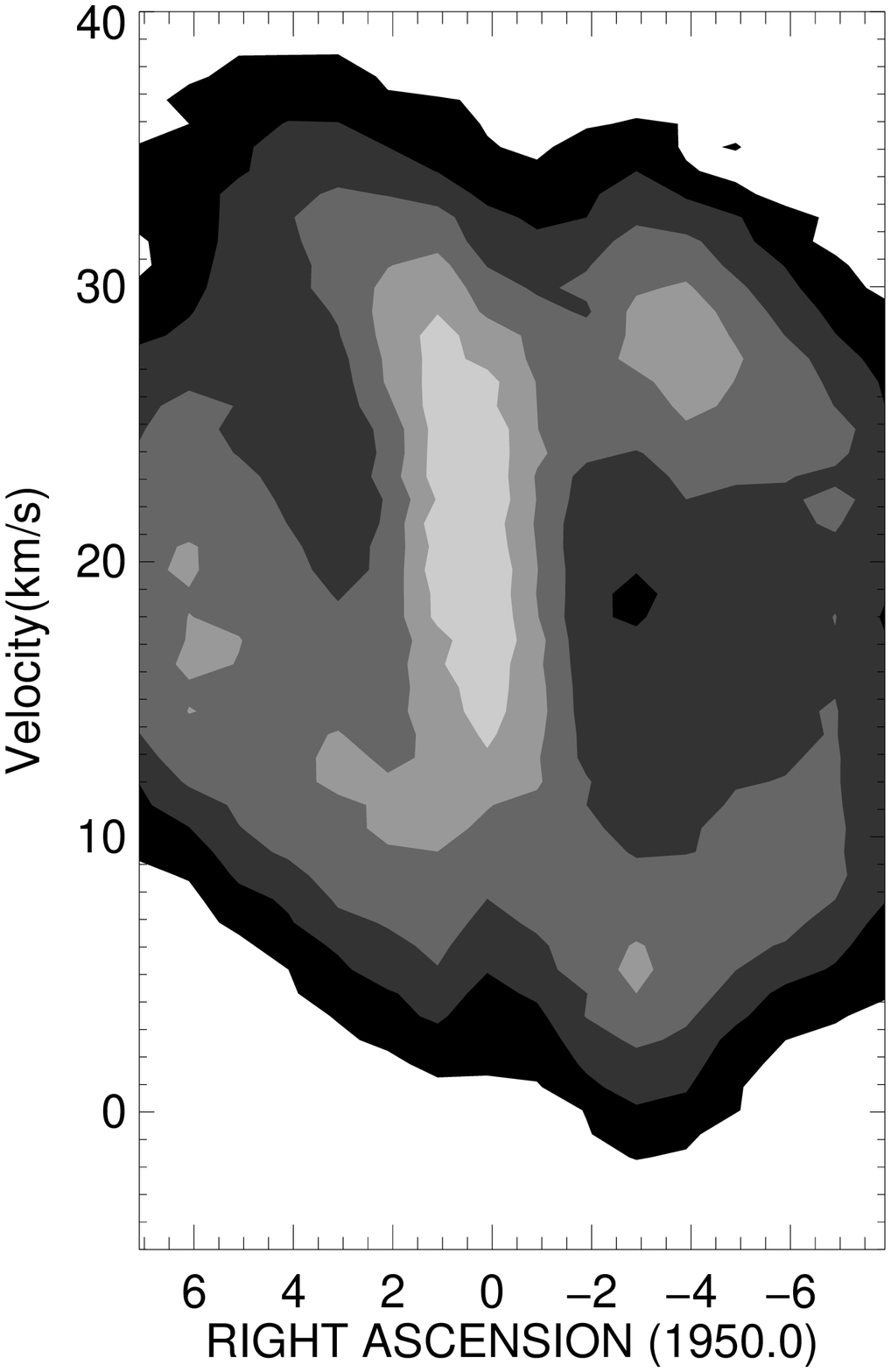] {Position-velocity diagram through the
strong arc of emission in Mon R2 IRS1. The contours 
are drawn at intervals of $1 \over 6$ of the peak value. 
\label{fig:monr2pv} }

\clearpage

\begin{figure}
\plotone{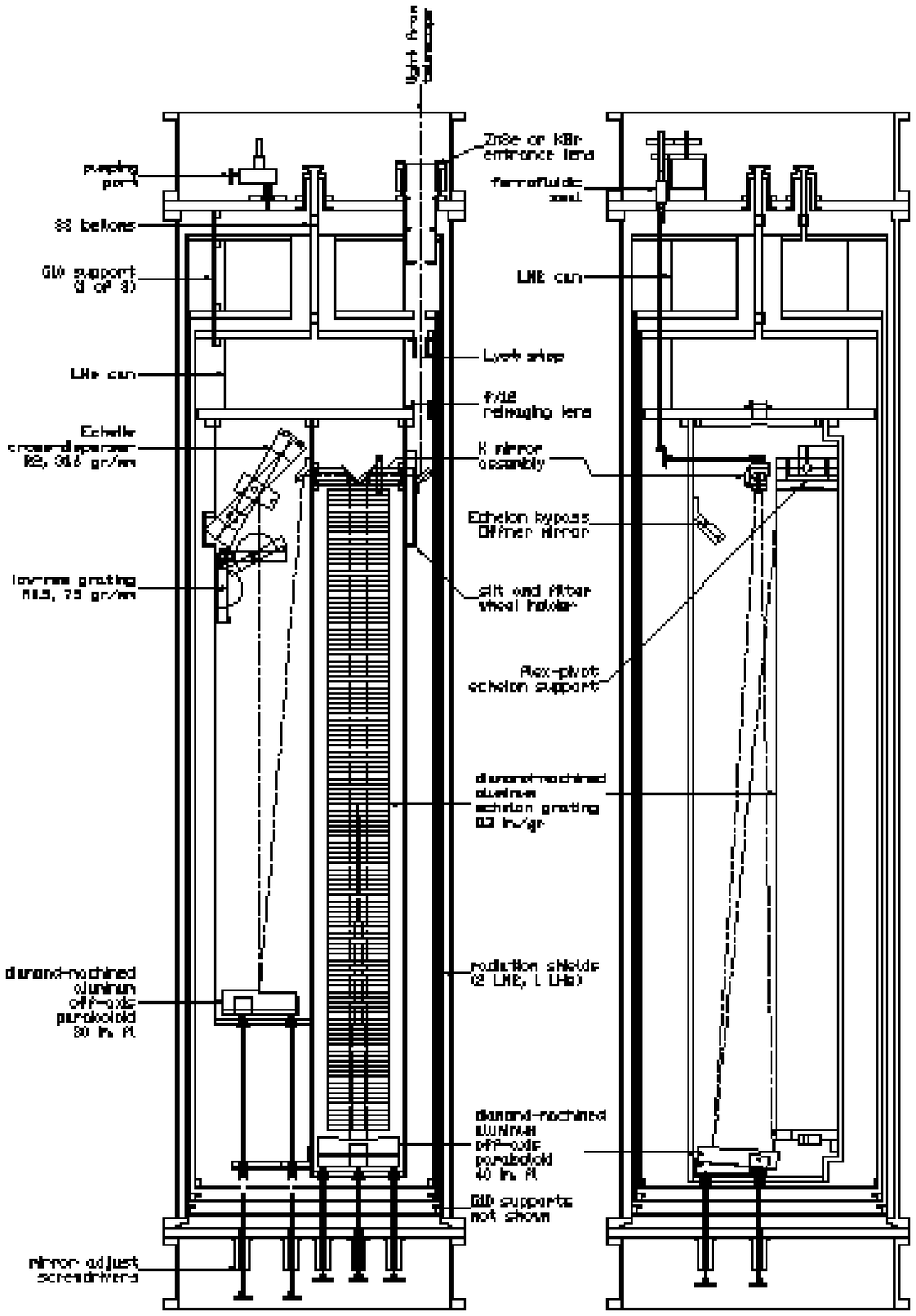}
\end{figure}

\begin{figure}
\plotone{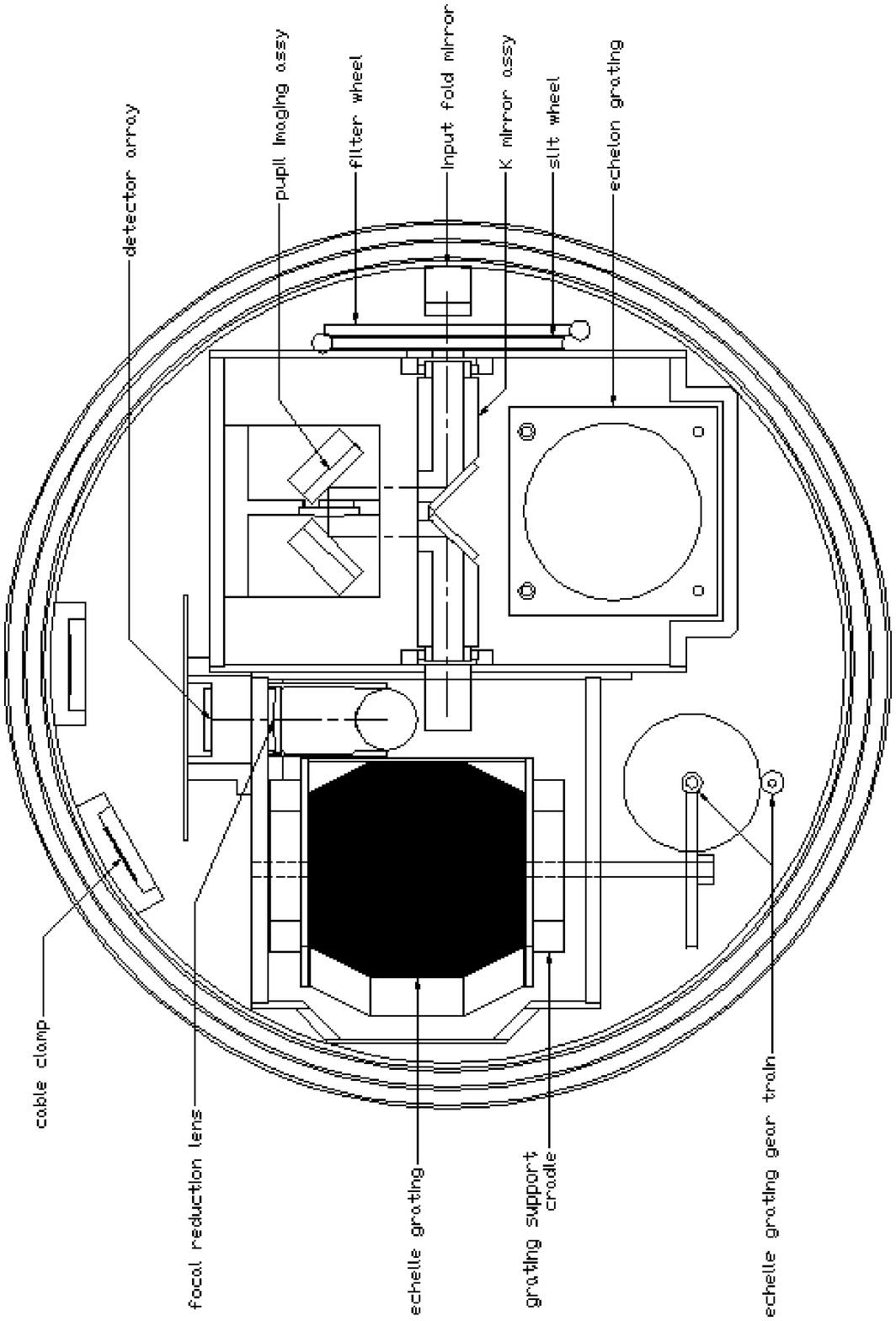}
\end{figure}

\begin{figure}
\plotone{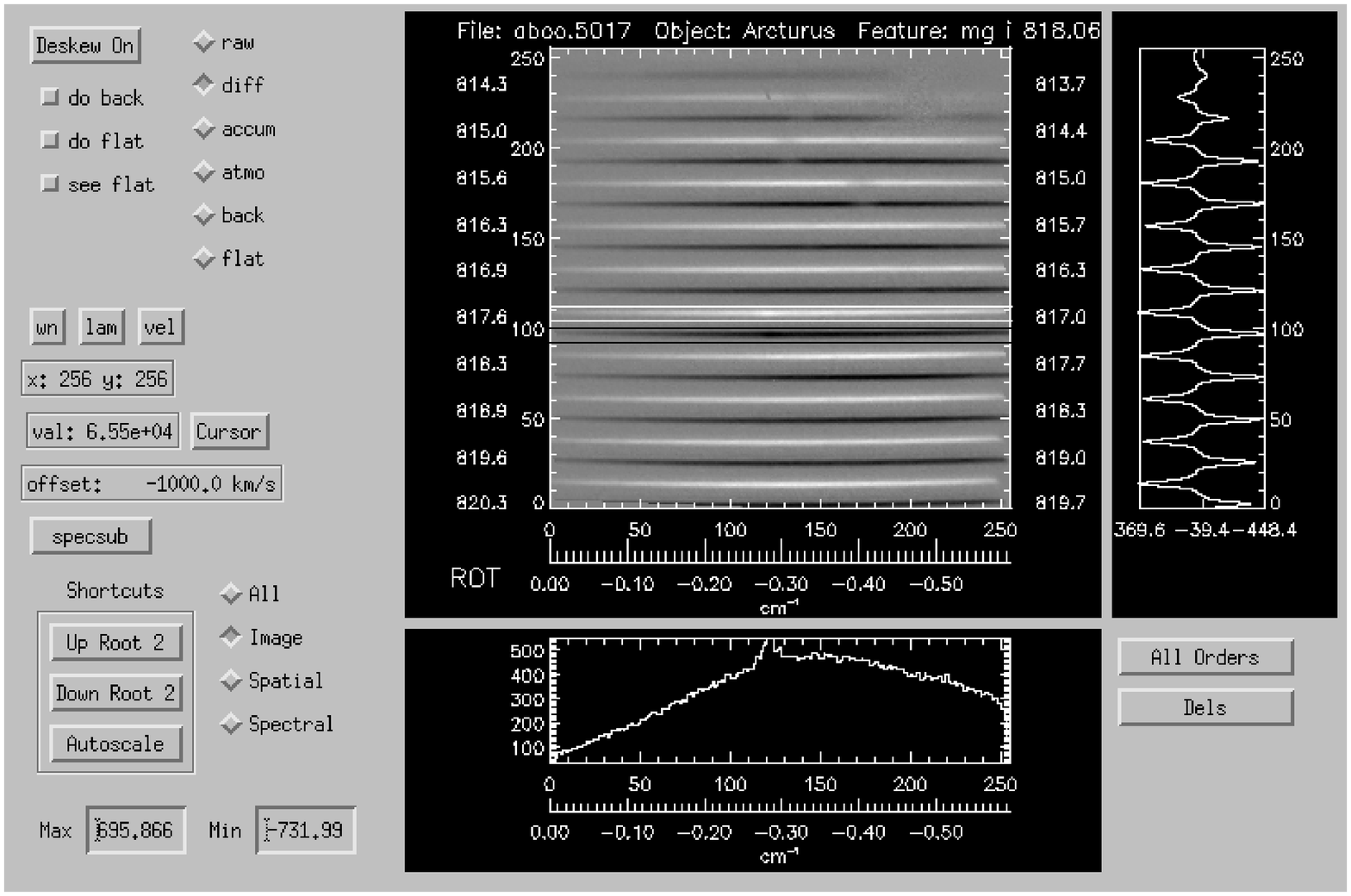}
\end{figure}

\begin{figure}
\plotone{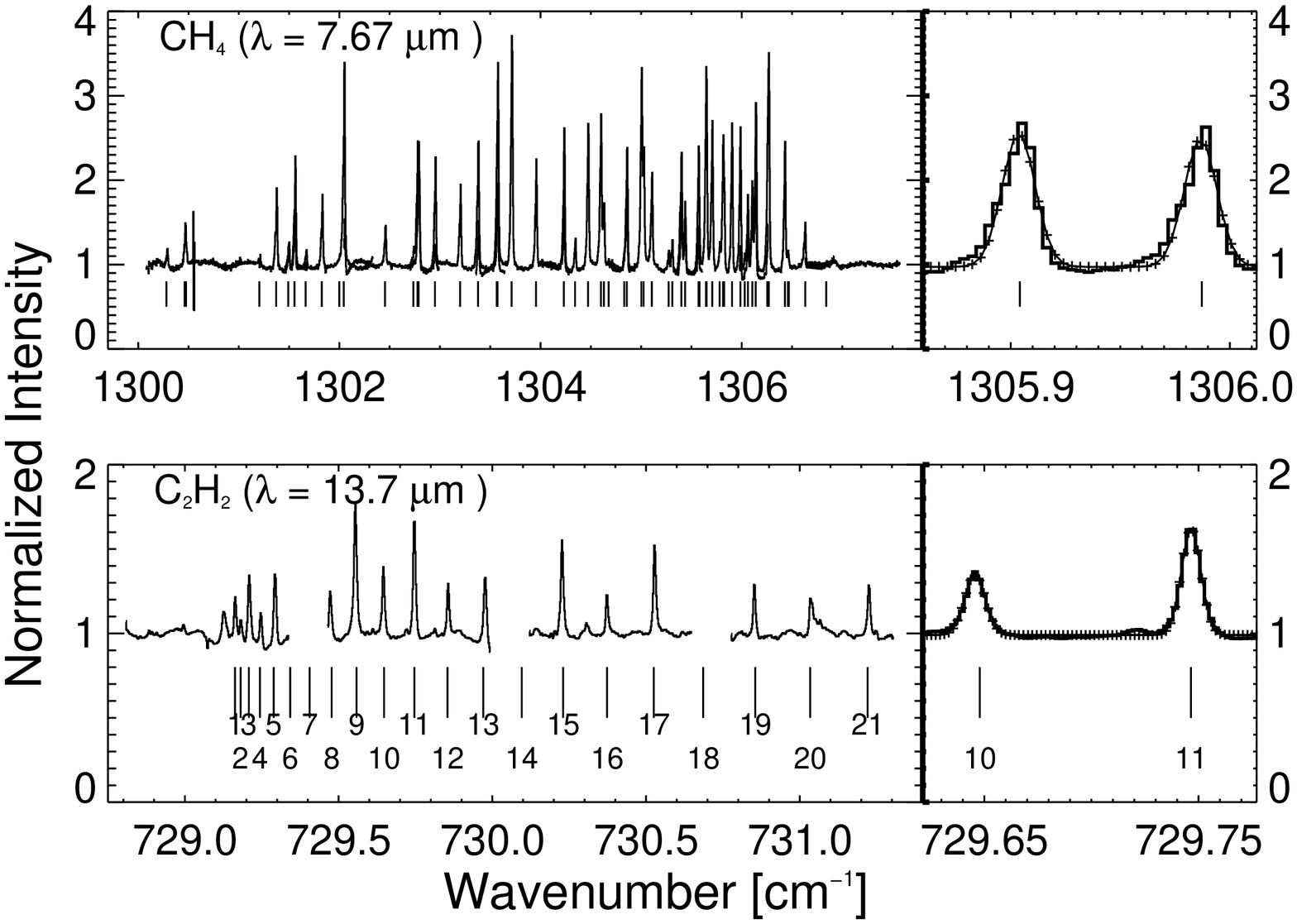}
\end{figure}

\begin{figure}
\plotone{fg5.eps}
\end{figure}

\begin{figure}
\plotone{fg6.eps}
\end{figure}

\begin{figure}
\plotone{fg7.eps}
\end{figure}

\begin{figure}
\plotone{fg8.eps}
\end{figure}

\begin{figure}
\plotone{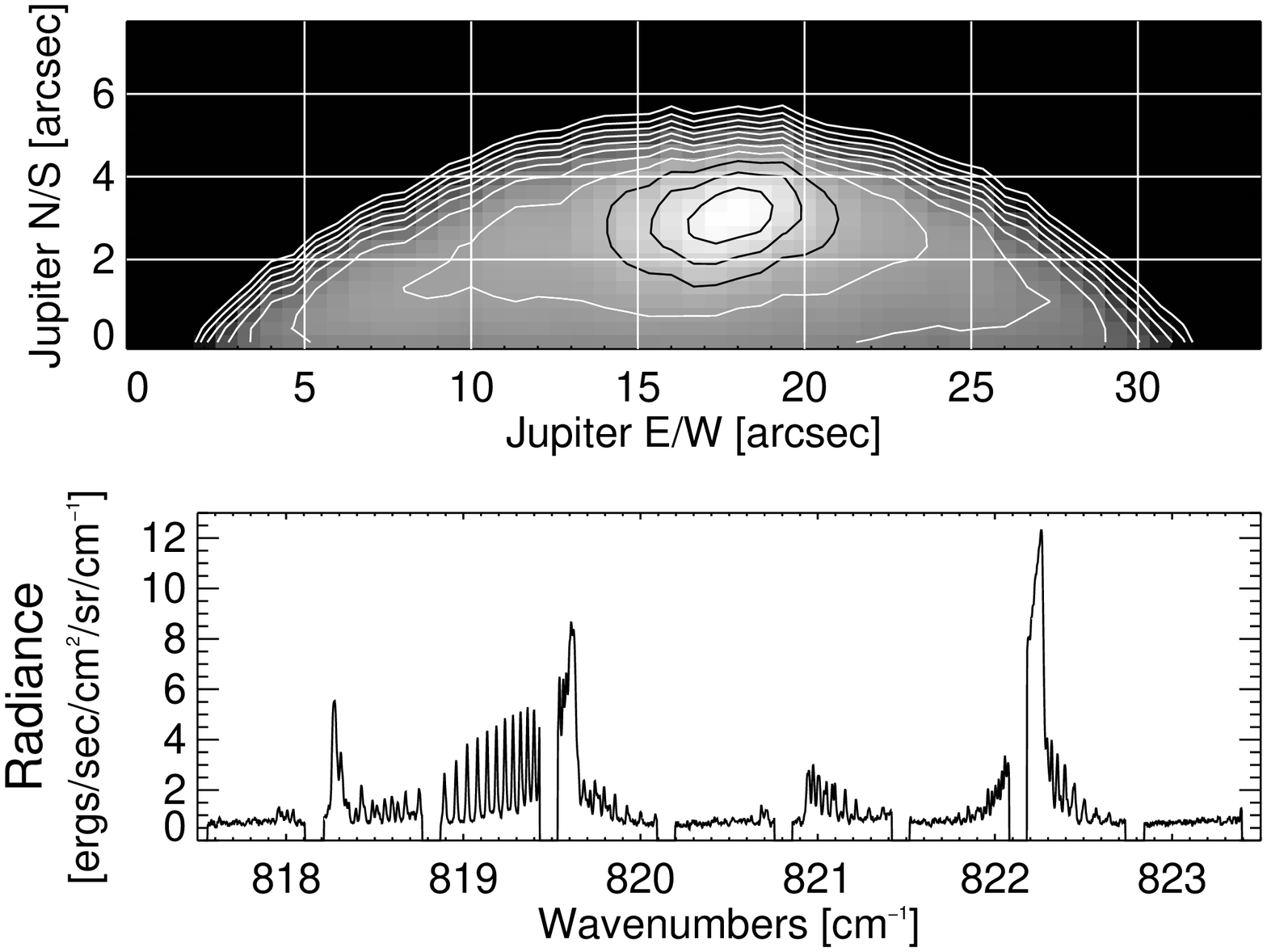}
\end{figure}

\begin{figure}
\plotone{fg10.eps}
\end{figure}

\begin{figure}
\plotone{fg11.eps}
\end{figure}

\end{document}